\documentclass[12pt,preprint]{aastex}                                         

\begin{document}
 
\title{Web Availability of MACHO Data}
 
\author{
    R.A.~Allsman\altaffilmark{1} and
    T.S.~Axelrod\altaffilmark{2}\\
      {For the MACHO Collaboration}
}
\altaffiltext{1}{Supercomputer Facility, Australian National University,
    Canberra, ACT 0200, Australia \\
    Email: {\tt robyn.allsman@anu.edu.au}}

\altaffiltext{2}{Research School of Astronomy and Astrophysics,
        Mount Stromlo Observatory, Cotter Road, Weston, ACT 2611, Australia\\
 Email: {\tt tsa@mso.anu.edu.au}}

\begin{abstract}
The MACHO Project generated two-color photometric lightcurves for 73
million stars in the LMC, SMC, and the galactic bulge during its 8
years of observing.  This photometry, along with all images and a
catalog of LMC variable stars, is now available for viewing or
download from the MACHO Project websites, http://wwwmacho.anu.edu.au
or http://wwwmacho.mcmaster.ca.  The capabilities and organization of
the new data retrieval facility is described in this paper.
\end{abstract}

\section{Introduction}
The MACHO Project \citep{alc00} had as its
principal goal the detection of gravitational microlensing in the LMC,
with the SMC and Galactic bulge as secondary targets.  The 1.3m Great 
Melbourne Telescope at Mt. Stromlo Observatory was used for the 
observations, with a ccd mosaic camera containing 8 2k x 2k chips
supplying images in two passbands simultaneously.  Observations
began in 1992, and terminated in 2000.  Approximately 8TB of image
data were collected, and photometric lightcurves constructed for
about $7.3 * 10^7$ stars.

All MACHO images and lightcurves, and a catalog of LMC periodic
variable stars, are now publicly available from the MACHO project
website,\\ 
http://wwwmacho.anu.edu.au/Data/MachoData.html or\\
http://wwwmacho.mcmaster.ca/Data/MachoData.html. 
The MACHO data archive is housed on a mass
storage system at the ANU Supercomputer Facility.  The storage
hardware is a  robotic tape silo, fronted by a large disk cache.
Access times vary from near zero if data is cached, to several
minutes if data must be fetched from tape.  Users should be aware that
the data they request will be delivered from the data archive in
Australia regardless of which web site they utilize.
  
The following sections of the paper briefly describe the data that
is now available and the nature of the web interface that is provided
to it.  More detailed information is available from the help sections
provided on the website.

\section{Image Access}

Images can be selected in several ways.  If only a representative
image of a field is required, the best choice is probably the ``Template 
Image Download'' interface.  Template images form the base of Macho
photometry and astrometry, and have been selected to have good seeing
and dark sky conditions.  The ``Macho Image Search'' interface provides
a flexible method of locating images of interest.  Selection criteria can
include position, date of observation, seeing, airmass, and sky
brightness.  Finally, the ``Image Download'' interface is for situations
where the desired observation id's are already known.  

\section{LightCurve Access}

The Macho photometry data is internally organized as lightcurves.  A
lightcurve contains all available data in both passbands for a given
star.  At each time point several quantities are available in addition
to the two instrumental magnitudes.  These include the estimated errors,
flags that indicate possible problems with the particular data point,
and information about the observation that generated the data.  

At present, the stars for which lightcurves are desired must be specified
by their Macho field.tile.sequence identifier.  This allows easy access
to lightcurves which have appeared in published papers, but is not useful
for searches, eg for objects near some particular position.  As discussed
in the last section of this paper, an enhanced search capability will
be provided shortly.

Once selected, lightcurves can either be viewed interactively, or
downloaded to the user's computer.  The interactive viewer provides
capabilities to zoom in or out, and to access details of individual
lightcurve points.

\section{Variable Star Catalog}

The variable star catalog contains about 21000 stars from the LMC.  These
stars have been phased and classified, and the catalog contains a variety
of quantities of interest.  Stars may be selected based on their catalog
parameters, and phased or unphased lightcurves displayed.  A paper 
presenting the variable star catalog is in preparation \citep{alc01}.

\section{Future Enhancements}

A number of enhancements to the Macho data access site are planned
for the near future:
\begin{itemize}
\item RA/Dec selection of stars:
The need to access lightcurves by field.tile.sequence will be removed with
a new interface that allows selection of stars based on RA/Dec coordinates.
\item GLU: 
A GLU interface will be provided, so that external websites, such as CDS,
will be able to easily incorporate Macho data.
\item Mirrors:
We plan to set up mirror sites in North America for the data, which is
currently available only from Australia.
\item Improved image format:
Image data is currently returned as 16 separate FITS files, one for each
amplifier in the CCD mosaic.  We are planning to bundle these into a single
multi-extension FITS file, and include WCS data in the header which will
allow any pixel in the image to be mapped to RA/Dec with sub-arcsecond
precision. 
\end{itemize}
\appendix

\end{document}